# Modeling Reflexivity of Social Systems in Disease Spread

**Minkyoung Kim**[1], **Raja Jurdak**[1], **Dean Paini**[2]

Data61, Commonwealth Scientific and Industrial Research Organisation (CSIRO), Australia[1]
Health & Biosecurity, CSIRO, Australia[2]
{Minkyoung.Kim, Raja.Jurdak, Dean.Paini}@csiro.au

**Abstract**

Diffusion processes in a social system are governed by external triggers and internal excitations via interactions between individuals over social networks. Underlying mechanisms are crucial to understand emergent phenomena in the real world and accordingly establish effective strategies. However, it is challenging to reveal the dynamics of a target diffusion process due to invisible causality between events and their time-evolving intensity. In this study, we propose the Latent Influence Point Process model (LIPP) by incorporating external heterogeneity and internal dynamics of meta-populations based on human mobility. Our proposed model quantifies the reflexivity of a social system, which is the level of feedback on event occurrences by its internal dynamics. As an exemplary case study, we investigate dengue outbreaks in Queensland, Australia during the last 15 years. We find that abrupt and steady growth of disease outbreaks relate to exogenous and endogenous influences respectively. Similar diffusion trends between regions reflect synchronous reflexivity of a regional social system, likely driven by human mobility.

## 1 Introduction

Diffusion processes in the real world often produce non-Poisson distributed event sequences, where inter-event times are highly clustered in a short term but separated by long term inactivity Barabasi (2005); Vázquez et al. (2006). Examples are observed in both human and natural activities such as resharing microblogs in online social media Cheng et al. (2016); Kim, Newth, and Christen (2013b); Zhao et al. (2015), citing scholarly publications Kim, McFarland, and Leskovec (2017); Kim, Newth, and Christen (2014); Shen et al. (2014); Sinatra et al. (2015), a high incidence of crime along hotspots Mohler et al. (2011); Short, Bertozzi, and Brantingham (2010), and aftershock sequences near the seismic center Stein (1999). These all imply that an event occurrence is likely triggered by preceding events in cascades of different scales, and the timing of discontinuous events conveys information of underlying diffusion mechanisms.

Uncovering such feedback mechanisms between preceding and triggered events has drawn significant attention from a wide range of scientific communities Albala-Bertrand (2000); Crane and Sornette (2008); Kim, Jurdak, and Paini (2017), since it helps predict diffusion trends and establish cost effective strategies for the promotion or restriction of the diffusion process. When it comes to epidemics, an accurate understanding of underlying dynamics is more demanding for the timely control of infectious disease spread. However, it is very challenging due to often unknown causal relationships between events and limited information of private social networks, unlike explicit citation relationships in online social media or in academic publications. Moreover, large international and domestic travel volumes have increased the uncertainty of infection pathways. Thus, the quantification of exogenous and endogenous effects is essential to overcome the challenges and understand emergent bursts of infections. However, this has not been highlighted in epidemics.

In this regard, we propose the Latent Influence Point Process model (LIPP) which quantifies the reflexivity of a heterogeneous social system in disease spread by considering macro-level internal dynamics between meta-populations. The proposed model incorporates three major counterbalancing components: (1) exogenous influence covering environmental heterogeneity of target regions, (2) endogenous influence by incorporating human mobility across multiple regions, and (3) time decay effect on disease spread. We simulate our model and generate synthetic data as the ground truth to evaluate the model performance of parameter recovery, which validates our interpretation of dynamics with real data. As a case study, we investigate dengue spread in Queensland, Australia during 15 years (2002–2016), provided by Queensland Health. We also analyze human mobility patterns by using three different types of travel data such as int'l/domestic visitor survey and geo-tagged Twitter data. We find that abrupt versus precursory growth of infections are likely attributed to exogeneity versus endogeneity, respectively. Similar diffusion trends between regions reflect synchronous feedback mechanisms of regional social systems. That is, the timing and size of event bursts in each region are comparable, which is likely driven by human mobility. Finally, faster feedback (rapid time decay) likely comes up with the sharp increase of self and mutual excitations by preceding events (abrupt growth of clustered events).

To the best of our knowledge, this study is the first to incorporate macro-level internal dynamics between meta-populations based on human mobility as topological pathways in a social system. In this way, our proposed model uncovers self and mutual excitation nature of disease spread across multiple regions with rich context. We expect that our proposed framework can be readily applied to multivariate stochastic processes without prior knowledge of causal relationships among discrete events. Also, our model can provide the reflexivity of a complex system in diffusion processes, showing the level of responsiveness to preceding events by its internal dynamics.

## 2 Related Work

The spread of infectious diseases leads to form event clusters in both space and time. As discussed earlier, such space-time clustering of discrete events have been widely observed in diverse research areas, such as aftershock sequences near the epicenter in seismology Ogata (1988), burglary sequences along hotspots in criminology Mohler et al. (2011), and citation sequences in online social media Cheng et al. (2016); Kim, Newth, and Christen (2016); Myers, Zhu, and Leskovec (2012) or in science Kim, McFarland, and Leskovec (2017); Kim, Newth, and Christen (2014); Shen et al. (2014).

Such spatiotemporal events are well realized by a point process due to its flexible consideration of lasting impact of bursty behaviors rather than a current snapshot Kim, McFarland, and Leskovec (2017), and thus it is widely used as a mathematical tool in diverse research areas Daley and Vere-Jones (2007); Snyder and Miller (2012). For understanding the origin of a burst, the interplay between external shock and internal dynamics in complex systems has also been of great interest across disciplines Kim, Jurdak, and Paini (2017); Myers, Zhu, and Leskovec (2012); Roehner, Sornette, and Andersen (2004). When it comes to epidemics, exotic virus importation via international visitors from endemic regions may trigger a disease outbreak Morse (2001); WHO (2009), which further spreads in cascades via social interactions. Accordingly, the quantification of such exogenous and endogenous effects on the disease spread helps obtain an accurate insight into underlying processes, which has been however relatively neglected in epidemic studies Bhatt et al. (2013); Colizza and Vespignani (2008); Pastor-Satorras et al. (2015).

In addition, disease outbreaks are nationwide, spanning local, inter-state, and inter-country transmissions Roth et al. (2014). This spatial expansion is enabled by human mobility Jurdak et al. (2015); Wesolowski et al. (2015); Wilder-Smith and Gubler (2008) which can reflect the topological heterogeneity of meta-populations at a macro level Barrat, Barthlemy, and Vespignani (2008); Colizza and Vespignani (2008); Kim, McFarland, and Leskovec (2017); Kim, Newth, and Christen (2013a). Moreover, advancements in social sensing technology Aggarwal and Abdelzaher (2013); Kim and Jurdak (2017) collecting human mobility patterns from mobile devices or social media records can overcome the limitation of collecting private social network structures.

In this context, we use a point process approach as a fundamental framework for disease spread, incorporate human mobility patterns as more realistic and contextual inputs to our proposed model, and quantify exogenous and endogenous effects on diffusion.

Table 1: Notation.

| Sym. | Description |
|---|---|
| $\boldsymbol{R}$ | a set of regions |
| $r_n$ | region where $n$-th infection has occurred, $r_n \in \boldsymbol{R}$ |
| $t_n$ | time of $n$-th infection, $t_n \in [0, T]$ |
| $N$ | total number of infections |
| $\boldsymbol{D}$ | whole history of infections, $\boldsymbol{D} = \{(t_n, r_n)\}_{n=1}^{N}$ |
| $T$ | an observation period |
| $\eta_r$ | intrinsic environmental risk of region $r$, $\eta_r > 0$ |
| $\xi_r$ | endogenous influence of region $r$, $\xi_r > 0$ |
| $\rho_r^k$ | directional human mobility from region $k$ to $r$ such that $\sum_{k \in \boldsymbol{R}} \rho_r^k = 1$ |
| $\varphi$ | parameter of temporal relaxation function, $\varphi > 0$ |
| $\lambda_r(t)$ | infection intensity of region $r$ at time $t$ |
| $\lambda_r^k(t)$ | infection intensity of region $r$ influenced by region $k$ at time $t$ |

## 3 Proposed Model

In this section, we propose a new diffusion framework which quantifies exogenous and endogenous dynamics in disease spread by incorporating external heterogeneity and internal dynamics of meta-populations at a macro level.

### 3.1 Background

We consider a Hawkes process Hawkes (1971) as our fundamental diffusion framework, since it is a non-Markovian extension of the Poisson process and thus realizes the clustering of events in the real world. The general univariate Hawkes process is defined with an intensity function,

$$\lambda(t) = \mu + \sum_{t_i < t} g(t - t_i) , \qquad (1)$$

where the first term $\mu$ represents a background intensity by external influence. The second term characterizes endogenous intensity reflecting weighted response between events. That is, the intensity of event occurrences is dependent on the history of preceding events.

As an exemplary scenario in the real world, we focus on disease spread. As shown in Figure 1(a), disease infections are represented as a single arrival process, which disregards infections by self and mutual excitations (intra- and inter-region disease transmission) in Figure 1(b). As discussed earlier, such cross-regional outbreaks are accelerated by human mobility. In this context, the objective of our framework is to model bursty behavior of disease outbreaks across meta-populations by incorporating human mobility as topological pathways in a social system.

### 3.2 Latent Influence Point Process

We now propose the Latent Influence Point Process model (LIPP) which incorporates the exogeneity and endogeneity of a social system as major components for realizing bursty diffusion processes in the real world. Based on inputs of a spatial and temporal event sequence and cross-regional human mobility, our model aims to quantify the reflexivity of a social system using inferred model parameters.

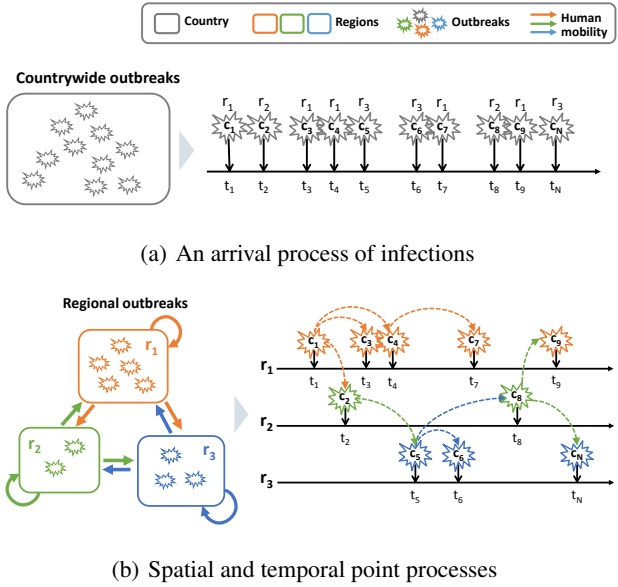

Figure 1: Disease spread as point processes. (a) countrywide outbreaks of an infectious disease over time ($c_i$: $i$-th contagion at time $t_i$ in region $r_j$, and $N$: total infections during an observation period). (b) countrywide outbreaks in (a) can be decomposed into timelines of each region, which can be represented as spatial and temporal point processes. Regions are color-coded, thick colored arrows represent human mobility between regions, and dashed arrows indicate hidden infection trajectories influenced by social interactions.

Suppose that we observe an event sequence $\boldsymbol{D}$ consisting of $N$ spatiotemporal events in a set of regions $\boldsymbol{R}$ during an observation time period $[0, T]$. Here, each event is represented by a pair of its occurrence time $0 < t_n < T$ and region $r_n \in \boldsymbol{R}$ as a tuple such that $\boldsymbol{D} = \{(t_n, r_n)\}_{n=1}^N$, and the events are sorted by their time moments.

As shown in Figure 1(b), we consider multiple timelines separated by event occurrence regions. For each region $r$, the history of events consists of two different types of event sequences, $\boldsymbol{D}_r^0$ and $\boldsymbol{D}_r^k$, influenced by external sources $0 \notin \boldsymbol{R}$ and triggered by preceding events at neighboring regions $k \in \boldsymbol{R}$, respectively. Given the whole event sequence $\boldsymbol{D}_r = \cup_{k \in \boldsymbol{R}_+} \boldsymbol{D}_r^k$, $\boldsymbol{R}_+ \equiv \boldsymbol{R} \cup \{0\}$, we assume that each event sequence is generated by a Poisson process. Thus, region $r$'s intensity function of time $\lambda_r(t)$ is defined based on the superposition property of Poisson processes Cinlar (2013) as

$$\lambda_r(t) = \lambda_r^0 + \sum_{k \in \boldsymbol{R}} \lambda_r^k(t) \ . \quad (2)$$

That is, we consider doubly stochastic point processes defined by $\lambda_r^k$ as our diffusion framework for the realization of intra- and inter-region disease transmissions, which corresponds to multi-dimensional Hawkes processes.

We incorporate three major counter-balancing components into our framework: (1) exogenous influence covering environmental heterogeneity of target regions, (2) endogenous influence attributed to macro-level interactions between meta-populations, and (3) time decay effect with an exponential memory kernel. Details are discussed as below.

**Exogenous Intensity.** In each region $r$, events can occur independently of the previous event history due to external influence. This is modeled with a Poisson process with a background intensity as

$$\lambda_r^0 = \eta_r \, \rho_r^0 \ , \quad (3)$$

The first term $\eta_r > 0$ denotes disease-specific environmental heterogeneity of region $r$ (environmental infectiousness of a target disease). That is, region $r$ has an intrinsic environmental risk that does not change much over time, such as average temperature and humidity, annual precipitation, and distributions of disease vectors (*e.g.*, mosquito vectors in dengue), and thus some area is more likely to cause disease outbreaks than others. The second term, $\rho_r^0 > 0$ represents the probability that an infection occurs in region $r$ by external exposures such that $\sum_{r \in \boldsymbol{R}} \rho_r^0 = 1$. For instance, international visitors from virus endemic regions outside $\boldsymbol{R}$ (*i.e.*, region 0) triggers local outbreaks in region $r$.

**Endogenous Intensity.** Contrary to exogenous infections, internal dynamics in a social system drives bursts of events going through interactions between individuals over social networks, so it is also called internal influence. Our model incorporates cross-regional human mobility as macro-level endogenous effects on diffusion, and the intensity brought by mutual excitations across multiple regions is defined as

$$\lambda_r^k(t) = \sum_{t_i < t} \xi_k \, \rho_r^k \, \phi(t - t_i) \ , \quad (4)$$

where $\xi_k > 0$ represents the latent influence (infectiousness) of region $k \in \boldsymbol{R}$ on other regions. For instance, the center of a city has the larger floating population than neighboring suburbs (connected by commute or travel routes), and thus it more likely triggers further infections in its neighboring regions. The second term $\rho_r^k > 0$ represents the strength of directed connectivity from region $k$ to $r$ based on human mobility patterns such that $\sum_{r \in \boldsymbol{R}} \rho_r^k = 1$. Finally, the third term captures the time relaxation function for reflecting the effect of time decay on the likelihood to spread. Here, we consider an exponential memory kernel such that $\phi(t - t_i) = \exp(-\varphi(t - t_i))$ where $\varphi > 0$ indicates the time decay parameter.

## 4 Inference

In this section, we infer our model parameters by using the Expectation-Maximization (EM) algorithm. As discussed in the previous model formulation, our model parameters are $\eta_r$, $\xi_r$, and $\varphi$ for each region $r \in \boldsymbol{R}$, representing environmental heterogeneity, latent influence, and time decay exponent, respectively.

**Likelihood.** According to the property that inter-event times generated by a Poisson process follow the exponential distributions, we first calculate the likelihood of observing an

event sequence $D = \{D_r\}_{r \in R} = \{(t_n, r_n)\}_{n=1}^N$ during an observation period $[0, T]$ as

$$p(D \mid \eta, \xi, \varphi) = \exp\left(-\int_0^T \sum_{r \in R} \lambda_r(t) dt\right) \prod_{i=1}^N \lambda_{r_i}(t_i) , \quad (5)$$

where $\eta = \{\eta_r\}_{r \in R}$ and $\xi = \{\xi_r\}_{r \in R}$ denote all target regions' environmental heterogeneity and latent influence respectively.

We introduce the latent index variables $Z = \{z_i\}_{i=1}^N$ consisting of event indicators each of which has triggered the $i$-th event, since causal relationships between events are unknown as discussed in Figure 1(b) (dashed lines). Each latent variable $z_i = \{z_{ij}\}_{j=0}^{i-1}$ is represented as an $i$-dimensional binary vector (e.g., $z_i = [0, 1, 0, ..., 0]^\top \in \mathbb{R}^i$ when $i$-th event is triggered by the second event). We impose a Bernoulli distribution prior as

$$p(z_i) = \prod_{j=0}^{i-1} (\pi_{ij})^{z_{ij}} , \quad \pi_i = \{\pi_{ij}\}_{j=0}^{i-1} , \quad (6)$$

where $\pi_{ij} > 0$, $\sum_{j=0}^{i-1} \pi_{ij} = 1$, and $\pi_i$ is a set of probabilities of selecting preceding event $j \in \{0, ..., i-1\}$ which has triggered event $i$. Note that $j$ starts from 0 representing external influence so that a prior covers an initial event.

The likelihood in Eq. (5) with the latent variables is

$$p(D, Z \mid \Theta, \Pi) = p(D \mid Z, \Theta) p(Z \mid \Pi)$$

$$= \exp(-\mathfrak{L}) \prod_{i=1}^N \prod_{j=0}^{i-1} \left(\lambda_{r_i}^{r_j}(t_i)\right)^{z_{ij}} \prod_{i=1}^N \prod_{j=0}^{i-1} (\pi_{ij})^{z_{ij}} , \quad (7)$$

where $\mathfrak{L} = \int_0^T \sum_{r \in R} \lambda_r(t) dt$, $\Theta = \{\eta, \xi, \varphi\}$, and $\Pi = \{\pi_i\}_{i=1}^N$. The log likelihood of Eq. (7) is represented as

$$\log p(D, Z \mid \Theta, \Pi)$$

$$= \sum_{i=1}^N \sum_{j=0}^{i-1} z_{ij} \left(\log \lambda_{r_i}^{r_j}(t_i) + \log \pi_{ij}\right) - \mathfrak{L} . \quad (8)$$

**Expectation-Maximization.** In order to find the maximum likelihood estimates with latent indices in Eq. (8), we apply the Expectation-Maximization algorithm.

In the E-Step, we calculate the expectation of the joint log likelihood in Eq. (8) given the parameters $\Theta$ and $\Pi$ as

$$\mathbb{E}_{Z \mid D, \Theta, \Pi}[\log p(D, Z \mid \Theta, \Pi)]$$

$$= \sum_{i=1}^N \sum_{j=0}^{i-1} \gamma(z_{ij}) \left(\log \lambda_{r_i}^{r_j}(t_i) + \log \pi_{ij}\right) - \mathfrak{L} , \quad (9)$$

where the responsibility is defined as

$$\gamma(z_{ij}) := \mathbb{E}_{Z \mid D, \Theta, \Pi}[z_{ij}]$$

$$= \frac{\pi_{ij} \lambda_{r_i}^{r_j}(t_i; t_j) \exp\left(-\int_{t_j}^{t_i} \lambda_{r_i}^{r_j}(t) dt\right)}{\sum_{j=0}^{i-1} \pi_{ij} \lambda_{r_i}^{r_j}(t_i; t_j) \exp\left(-\int_{t_j}^{t_i} \lambda_{r_i}^{r_j}(t) dt\right)}. \quad (10)$$

In the M-Step, on the other hand, we find the optimal parameters which maximize the expectation of the log likelihood in Eq. (9). We apply gradient descent for $\Theta$ and $\Pi$. For more details, see Appendix.

---

**ALGORITHM 1:** Simulation of LIPP

**Input** : $|R|, \eta_r, \rho_r^k, \xi_r, \varphi, T$ $(1 \leq r \leq |R|, 0 \leq k \leq |R|)$
**Output:** $D$

1. $D \leftarrow \varnothing$
2. $t_0 \leftarrow 0$
3. $i \leftarrow 0$
4. **while** $t_i < T$ **do**
5.     $i \leftarrow i + 1$
6.     $D_i \leftarrow \varnothing$
7.     $u \leftarrow U(0, 1)$
8.     **for** $r = 1, ..., |R|$ **do**
9.         $\Delta t_{r,i}^0 \leftarrow -\frac{1}{\varphi} \log(1 - u)$
10.         $D_i \leftarrow D_i \cup \Delta t_{r,i}^0$
11.         **for** $k = 1, ..., |R|$ **do**
12.             $\Delta t_{r,i}^k \leftarrow -\frac{1}{\varphi} \log\left(1 + \frac{\varphi}{\lambda_r^k(t_{i-1})} \log(1 - u)\right)$
13.             $D_i \leftarrow D_i \cup \Delta t_{r,i}^k$
14.     $t_i \leftarrow t_{i-1} + \min(D_i)$
15.     $(r^*, k^*) = \arg\min_{r,k} D_i$
16.     $D \leftarrow D \cup t_i$
17.     **for** $r = 1, ..., |R|$ **do**
18.         **for** $k = 1, ..., |R|$ **do**
19.             $\lambda_r^k(t_i) \leftarrow \lambda_r^k(t_{i-1}) e^{-\varphi(t_i - t_{i-1})} + \xi_{r^*} \rho_r^{r^*}$

## 5 Simulation

We simulate our proposed model so that we can generate synthetic data as the ground truth and evaluate the model performance. As evaluation metrics, parameter recovery errors are examined with respect to the mean absolute percentage error (MAPE). We also evaluate relative strengths between estimated parameters, which is important to validate the interpretations of underlying diffusion processes with real data. After verifying the proposed model performance with synthetic data, we conduct experiments on real data from epidemics and interpret the dynamics of disease spread as a case study in the next section.

### 5.1 Synthetic Data Generation

We describe how to simulate our model LIPP with given parameters. With generated synthetic data, we verify our model by examining how well the model parameters are recovered in terms of parameter value and rank.

**Simulation Algorithm.** Our proposed model is the multi-dimensional Hawkes process, where events are mutually excited across more than one regions. Based on the doubly stochastic property of LIPP, the probability that a new event $i$ arrives within a time interval $\tau$ should consider the whole likelihood of that event generated by the intensity of mutual excitation, $\lambda_r^k(t)$ from an affecting region $k \in R_+$ ($R_+ \equiv R \cup \{0\}$) to a target region $r \in R$ as

$$F_{\Delta t_i}(\tau) = P(\Delta t_i < \tau) = \prod_{r \in R} \prod_{k \in R_+} F_{\Delta t_{r,i}^k}(\tau) , \quad (11)$$

where $F_{\Delta t_{r,i}^k}(\tau)$ is the CDF of inter-event time $\Delta t_{r,i}^k = t_{r,i}^k - t_{r,i-1}^k$ produced by the intensity $\lambda_r^k(t)$.

For exogenous infection ($k = 0$), the CDF of its inter-event time $\Delta t^0_{r,i}$ follows

$$F_{\Delta t^0_{r,i}}(\tau) = 1 - \exp\left(\eta_r \, \rho^0_r \, \tau\right) \quad . \quad (12)$$

On the other hand, for the simulation of mutual excitations ($k \neq 0$), we represent the CDF of $\Delta t^k_{r,i}$ as

$$F_{\Delta t^k_{r,i}}(\tau) = 1 - \exp\left(-\frac{1}{\gamma} \lambda^k_r(t_{i-1})\left(1 - e^{-\gamma\tau}\right)\right) \quad . \quad (13)$$

We then simulate LIPP by sampling $u$ from a uniform distribution $U(0,1)$ and find the corresponding elapsed time $\Delta t^0_{r,i}$ and $\Delta t^k_{r,i}$ using the inverse transform method Knuth (1998),

$$\Delta t^0_{r,i} = -\frac{1}{\gamma}\log(1-u) \quad , \quad (14)$$

$$\Delta t^k_{r,i} = -\frac{1}{\gamma}\log\left(1 + \frac{\gamma}{\lambda^k_r(t_{i-1})}\log(1-u)\right) \quad , \quad (15)$$

where $\lambda^k_r(t_i) = \lambda^k_r(t_{i-1})e^{-\gamma\Delta t_i} + \xi_k \rho^k_r$ by which we update the intensity for each iteration.

Now, $\Delta t_i$ can be chosen with the minimum inter-arrival time among all $(r \times (k+1))$ samples generated by exogenous and endogenous intensities as

$$\Delta t_i = \min\{\Delta t^0_{r,i}, \Delta t^k_{r,i}\}_{r \in \mathbf{R}, k \in \mathbf{R}+} \quad . \quad (16)$$

By sampling both $\Delta t^0_{r,i}$ and $\Delta t^k_{r,i}$ and choosing the minimum $\Delta t_i$, we finally can obtain information about the triggering source of event $i$. This procedure is described in Algorithm 1, and its output is a sequence of timestamps across all regions $\mathbf{R}$.

**Parameter Setting.** We generate 3 groups of synthetic datasets by varying the number of regions such that $|\mathbf{R}| \in \{5, 10, 15\}$, which imitates the number of target regions in our real data. Detailed descriptions on real data will be discussed in the next section. By replicating the estimated parameter values with our real data, we set parameter values for each group such as region $r$'s environmental heterogeneity $\eta_r \in [0.01, 1]$ and latent influence $\xi_r \in [0.5, 2]$. Each group include three different memory kernel exponents such that $\varphi \in \{1, 2, 3\}$, and thus there are 9 subgroups (3 groups × 3 exponents). For each subgroup, we generate 30 cases with random values of $\eta_r$ and $\xi_r$ within the specified ranges above. In total, there are 270 event histories for the synthetic data (9 subgroups × 30 test cases).

### 5.2 Experiments on Synthetic Data

In this section, we verify our proposed model with the simulation data based on parameter recovery errors.

**Parameter Value Recovery.** In order to examine how well our model parameters are recovered from data, we conducted experiments on the three groups of synthetic datasets. Due to the different scale of each parameter, we employ the mean absolute percentage error (MAPE) as a metric for evaluating parameter recovery accuracy as

$$\text{MAPE} = \frac{1}{n}\sum_{i=1}^{n}\left|\frac{A_i - F_i}{A_i}\right| \quad , \quad (17)$$

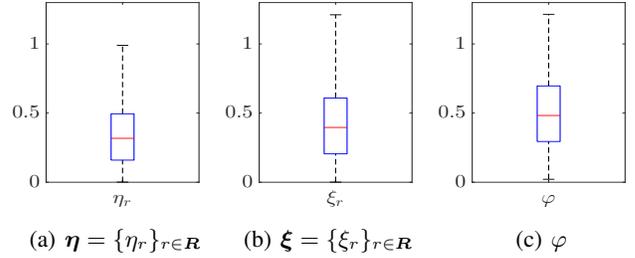

(a) $\boldsymbol{\eta} = \{\eta_r\}_{r \in \mathbf{R}}$    (b) $\boldsymbol{\xi} = \{\xi_r\}_{r \in \mathbf{R}}$    (c) $\varphi$

Figure 2: Box plots for the mean absolute percentage errors (MAPE) of estimated parameter values: (a) region $r$'s environmental heterogeneity $\eta_r$, (b) latent influence $\xi_r$, and (c) time decay parameter $\varphi$.

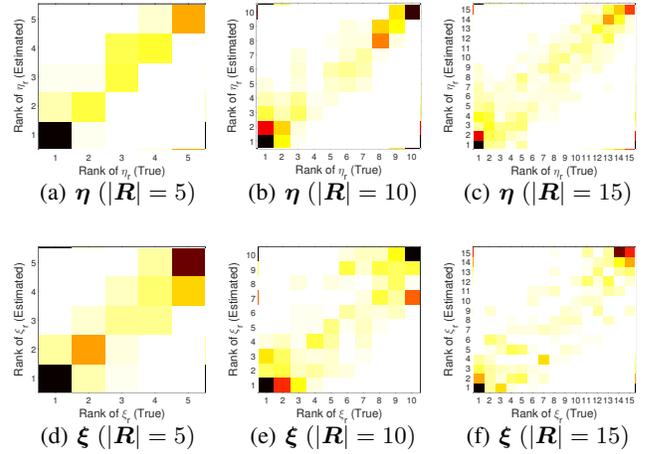

(a) $\boldsymbol{\eta}$ ($|\mathbf{R}| = 5$)   (b) $\boldsymbol{\eta}$ ($|\mathbf{R}| = 10$)   (c) $\boldsymbol{\eta}$ ($|\mathbf{R}| = 15$)

(d) $\boldsymbol{\xi}$ ($|\mathbf{R}| = 5$)   (e) $\boldsymbol{\xi}$ ($|\mathbf{R}| = 10$)   (f) $\boldsymbol{\xi}$ ($|\mathbf{R}| = 15$)

Figure 3: Relative strengths of estimated parameter values compared with the ground truth: (a) to (c) for region $r$'s environmental heterogeneity $\eta_r$ and (d) to (e) for latent influence $\xi_r$. In each plot, $x-$ and $y-$axes represent the pair of true and estimated ranks by varying the number of regions, $|\mathbf{R}| \in \{5, 10, 15\}$. The frequency of pairs is color-coded (darker color indicates higher frequency). Shaded grids along the line of $y = x$ imply that the parameter rankings are well recovered.

where $A_i$ and $F_i$ represent the actual and forecast parameter values for $n$ synthetic datasets.

Figure 2 depicts the box plot for the estimated parameter errors in MAPE, where parameter values are recovered with about a 40% error on average. That is, our proposed model recovers parameter values with a difference less than one magnitude order compared to the ground truth.

**Parameter Rank Recovery.** However, MAPE cannot show the recovery of relative strengths of parameter values. When it comes to epidemics, the rankings of estimated regional influences are crucial to establish effective strategies. In this regard, we also examine the recovery of relative strengths of true environmental heterogeneity $\boldsymbol{\eta}$ (Figure 3(a)-(c)) and endogenous influence $\boldsymbol{\xi}$ (Figure 3(d)-(f)) for each group ($|\mathbf{R}| \in \{5, 10, 15\}$) by ranking the true ($x$-axis) and its estimated ($y$-axis) parameter values in descending order. Each group has 3 subgroups ($\varphi \in \{1, 2, 3\}$) consisting of 30 test cases, and thus there are 450, 900, and 1,450 rank pairs of true and estimated $\boldsymbol{\eta}$ and $\boldsymbol{\xi}$ for the groups $|\boldsymbol{R}| \in \{5, 10, 15\}$, respectively. In each plot, the darkness of a grid is proportional to the frequency of the pairs. As the figure shows, shaded grids are overall observed along the line of $y = x$, which shows the order of regional criticality is well recovered. More importantly, the rankings of the most and least influential regions are better recovered, which helps timely and cost effective control on disease spread.

Table 2: Data description of the International Visitor Survey (IVS), National Visitor Survey (NVS), and Twitter.

| Data | Collection Period | #Visits | #Persons |
| --- | --- | --- | --- |
| IVS | 2005 - 2015 | 1,274,903 | 442,445 |
| NVS | 1998 - 2015 | 751,184 | 588,323 |
| Twitter | 2015 | 925,945 | 79,271 |

## 6 Case Study: Dengue Spread

In this section, we conduct experiments on real data and interpret estimated model parameters based on the verification of parameter recovery with synthetic data in the previous section. We first describe the collection and preprocessing of real data and analyze experimental results.

### 6.1 Data Description

**Dengue Outbreaks.** We investigate dengue outbreaks in Queensland, Australia from 2002 to 2016, provided by Queensland Health. The dengue virus is a mosquito-borne viral disease transmitted among humans by mosquito vectors, whose outbreak risk is rapidly increasing worldwide WHO (2009). The data contains records of anonymised infected individuals such as onset dates, residence postcodes, and acquisition places if available. For understanding cross-regional infections at a macro level, we categorized residence postcodes into 15 regions which correspond to the statistical areal level 4 (SA4) defined by the Australian Statistical Geography Standard (ASGS). Based on selected target regions, we create an event sequence as a tuple consisting of occurrence time and region identity, i.e., $\boldsymbol{D} = \{(t_n, r_n)\}_{n=1}^N$.

**Human Mobility.** As discussed in the proposed model, we incorporate human mobility as topological heterogeneity across multiple regions, which reflects macro-level internal dynamics in a social system. In order to obtain structural connectivity between regions, we employ three different types of travel datasets such as International Visitor Survey (IVS), National Visitor Survey (NVS), and Twitter, as described in Table 2. IVS TRA (2017a) and NVS TRA (2017b) are conducted by Computer Assisted Personal Interviewing (CAPI) in the departure lounges of the international airports in Australia. These datasets consist of traveler information (e.g., home country or residence, gender, age group), travel time periods, trajectories of visiting locations in Australia. Finally, geo-tagged Twitter data provides daily movement patterns in contrast to irregular movements of tourists. Thus, combining these different types of human mobility datasets enables us to construct more exhaustive and complementary population fluxes across regions. We then construct a network whose nodes and links correspond to regions in a SA4 level and normalized fluxes between regions. The constructed network is accordingly shown in Figure 4, where the size of nodes represents external fluxes ($\rho_r^0$ in Eq. (3)) from outside of Queensland to each region and the width of links is proportional to internal fluxes ($\rho_r^k$ in Eq. (4)) between regions.

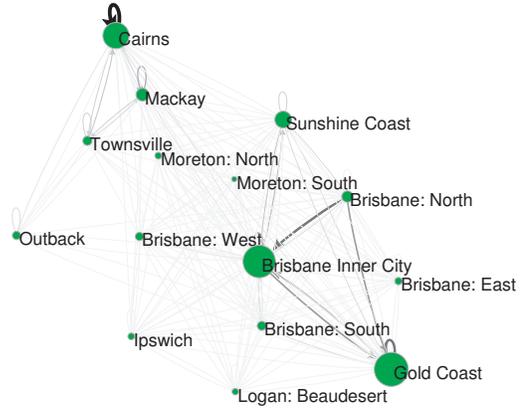

Figure 4: Estimated human mobility by using travel data in Table 2. The node size and link width are proportional to external fluxes from outside of Queensland and internal fluxes between regions, respectively. Self-loops and straight links represent intra- and inter-region human mobility.

### 6.2 Reflexivity of a Regional Social System in Disease Spread

The Hawkes process generalizes the nonhomogeneous Poisson process by allowing the self-exciting nature via preceding events, as discussed in Eq. (1). The linearity of the conditional intensity $\lambda(t)$ helps quantify the level of exogeneity and enogeneity in diffusion processes and align with a branching process consisting of triggers and descendants Daley and Vere-Jones (2007).

The branching ratio $b$ represents the average number of triggered events per initiating event and is defined as

$$b = \int_0^\infty g(t) dt \ . \quad (18)$$

**Reflexivity of Regional Social Systems.** Our proposed model extends multi-dimensional Hawkes process. From each regional intensity $\lambda_r(t)$ in Eq. (2), the background intensity and branching ratio correspond to $\mu = \eta_r \rho_r^0$ and $b = \frac{1}{\varphi} \sum_{k \in \boldsymbol{R}} \xi_k \rho_r^k$. Accordingly, we quantify the level of exogeneity $\mu$ and endogeneity $b$ for all target regions based on the estimated parameter values with our real data, as shown in Figure 5. For this quantification, we set the length of observation time window as one year in order to examine time-evolving internal dynamics with a fine-grained time resolution.

**Abrupt vs. Precursory Growth.** As Figure 5 shows, in general the background intensity $\mu$ hardly changes except for

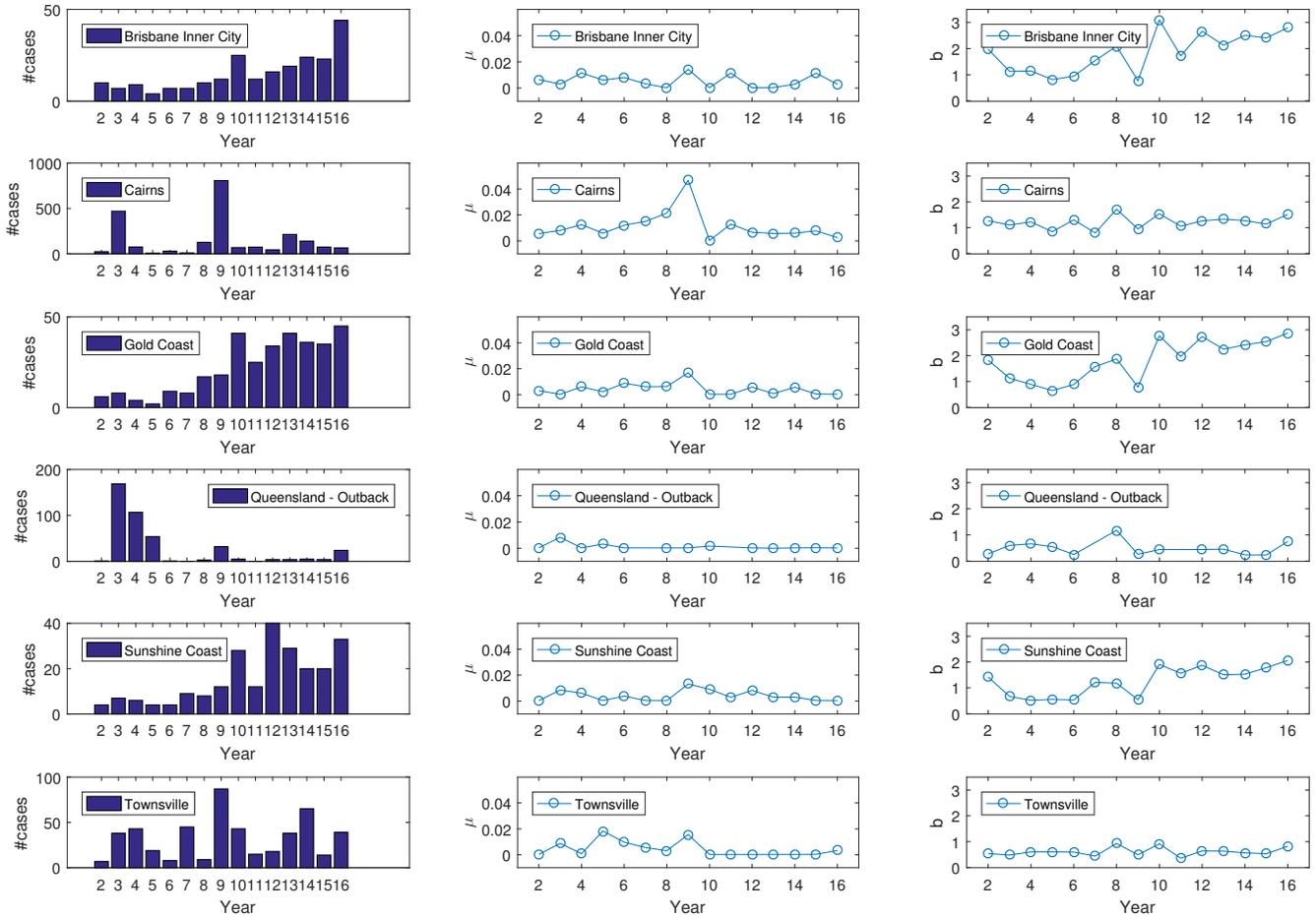

Figure 5: Reflexivity of regional social systems. The plots in the first column show the distributions of dengue outbreaks in each region during our observation period, while the second and third columns illustrate the level of exogeneity $\mu$ and endogeneity $b$ in dengue spread over time. Six regions are selectively chosen with the largest number of cases among 15 regions.

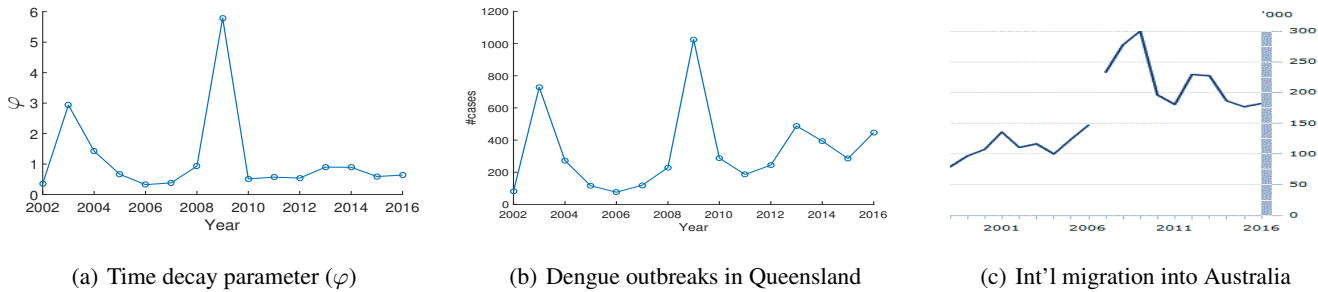

(a) Time decay parameter ($\varphi$)  (b) Dengue outbreaks in Queensland  (c) Int'l migration into Australia

Figure 6: (a) Estimation of time decay parameter $\varphi$ for each year and comparable population growth trends from the distributions of (b) dengue cases in Queensland and (c) overseas migration into Australia provided by the Australian Bureau of Statistics.

Cairns, while the branching ratio $b$ increases over time in metropolitan or populated ares such as Brisbane Inner City (BIC), Gold Coast (GC), and Sunshine Coast (SC). The exogeneity $\mu$ is the highest in 2009, and since then the endogeneity has greatly increased across regions. The first column in the figure exhibits two distinct diffusion patterns such as abrupt uprise in Cairns, Outback, and Townsville in 2003 and 2009 versus precursory growth in BIC, GC, and SC in 2010. The former cases in common show the higher exogeneity $\mu$ but with lower endogeneity $b$ in that period than others, while the latter shows the opposite patterns. That is, sharp growth is more likely attributed to ex-ternal influences, while gradual growth is likely due to internal dynamics via interactions between individuals in a social system. Also, the former and latter cases are from remote/periphery and metropolitan/populous areas respectively, showing a split in behavior.

**Synchronous Feedback Mechanisms.** Interestingly, BIC and GC exhibit the similar growth patterns and reflexivity of a regional social system, which implies that endogenous feedback mechanisms are synchronous. This is likely attributed to human mobility patterns, as shown in Figure 4. In other words, the large volumes of external visitors to these regions as well as large reciprocal fluxes between them more likely drive self and mutual excitations. Our proposed model can capture such emergent phenomena by incorporating endogenous effects of cross-regional human mobility on diffusion processes.

**Rapid Time Decay and Abrupt Growth.** Figure 6 shows (a) the estimated time decay parameter $\varphi$ in Eq. (4), (b) whole dengue outbreaks in Queensland during our data period, and (c) the growth of overseas migration into Australia provided by the Australian Bureau of Statistics (ABS). Interestingly, they show the similar curves of rises and falls particularly with the abrupt growth in 2009. In terms of time decay, fast feedback (rapid time decay) comes up with the sharp increase of self and mutual excitations by preceding events (abrupt growth of clustered events). This may have been affected by external influences such as the abrupt growth of international migrations into Australia in 2009 as human mobility is dynamic and nationwide.

# 7 Conclusion

We proposed a new model, LIPP which generalizes multi-dimensional Hawkes processes by incorporating macro-level internal dynamics between meta-populations, driven by human mobility. By introducing latent indicator variables for triggering events, our proposed model quantifies the level of exogeneity and endogeneity in disease spread without the need to know prior knowledge of causal relationships between events. These aspects increase the applicability of our proposed model to a wide range of intra- and inter-group diffusion processes in social and natural systems.

## A  Appendix

We explain how to find the optimal parameters $\Theta$ and $\Pi$ which maximize the expectation of the joint log likelihood, $\mathbb{E}$ in Eq. (9). First, we apply gradient descent to find the optimal solution for $\Theta = \{\boldsymbol{\eta}, \boldsymbol{\xi}, \varphi\}$ with the partial derivatives,

$$\frac{\partial \mathbb{E}}{\partial \eta_r} = \sum_{i=1}^{N} \delta(r_i, r) \frac{\gamma(z_i^0)}{\eta_r} - T\rho_r^0 \ , \tag{19}$$

$$\frac{\partial \mathbb{E}}{\partial \xi_r} = \sum_{i=1}^{N} \sum_{j=1}^{i-1} \delta(r_j, r) \frac{\gamma(z_{ij})}{\xi_r} - \frac{1}{\varphi} \sum_{i=1}^{N} \delta(r_i, r) (1 - \kappa) \ , \tag{20}$$

$$\frac{\partial \mathbb{E}}{\partial \varphi} = -\sum_{i=1}^{N} \sum_{j=1}^{i-1} \gamma(z_{ij})(t_i - t_j) + \frac{1}{\varphi^2} \sum_{i=1}^{N} \xi_{r_i} (1 - \kappa) - \frac{1}{\varphi} \sum_{i=1}^{N} \xi_{r_i} (T - t_i)\kappa \ , \tag{21}$$

where $\kappa := \exp(-\varphi(T - t_i))$.

On the other hand, we introduce Lagrangian multipliers $\lambda_i$ to find the optimal $\Pi$ as

$$\mathbb{E}' = \mathbb{E} - \sum_{i=1}^{N} \lambda_i \left( \sum_{j=0}^{i-1} \pi_{ij} - 1 \right) \ . \tag{22}$$

Now, the optimal solution for $\Pi$ should satisfy

$$\frac{\partial \mathbb{E}'}{\partial \pi_{ij}} = \frac{\gamma(z_{ij})}{\pi_{ij}} - \lambda_i = 0 \ , \tag{23}$$

$$\frac{\partial \mathbb{E}'}{\partial \lambda_i} = \sum_{j=0}^{i-1} \pi_j - 1 = 0 \ . \tag{24}$$

Therefore,

$$\pi_{ij} = \frac{\gamma(z_{ij})}{\sum_{j=0}^{i-1} \gamma(z_{ij})} \ . \tag{25}$$